\documentstyle[11pt,epsfig]{article}
\renewcommand{\baselinestretch}{2}

\begin{document}
\title{Electronic properties of germanene nanoribbons in external fields \\}
\author{M. H. Lee$^{a,b}$, J. Y. Wu$^{b}$, S. C. Chen$^{a}$, C. W. Chiu$^{c*}$
 and  M. F. Lin$^{a*}$ \\
\small ${^a}$Department of Physics, National Cheng Kung University,
 Tainan, Taiwan\\
\small $^b$ Center of General Studies, National Kaohsiung Marine University,
 Kaoshiung, Taiwan\\
\small ${^c}$ Department of Physics, National Kaohsiung Normal University,
Kaohsiung, Taiwan\\
}

\maketitle



\renewcommand{\baselinestretch}{2}
\maketitle
\begin{abstract}
Germanene nanoribbons, with buckled structures, exhibit unique electronic properties. The complicated relations among the quantum confinement, the spin-orbital coupling, the magnetic quantization, and the electric field dominate quantum numbers, energy dispersions, energy gap, state degeneracy, and wave functions. Such mechanisms can diversify spatial charge distributions and spin configurations on distinct sublattices. There exist the spin-split quasi-Landau levels and the valley-dependent asymmetric energy spectrum in a composite electric and magnetic field, manly owing to the destruction of z=0 mirror symmetry. The rich electronic structures are revealed in density of states as a lot of special structures. The predicted results could be directly verified by the scanning tunneling spectroscopy.

\end{abstract}

\par\noindent ~~~~$^*$Corresponding authors- E-mail: giorgio@fonran.com.tw (C. W. Chiu),\\E-mail: mflin@mail.ncku.edu.tw (M. F. Lin)
\\

\vskip0.6 truecm

$\mathit{PACS}$: 75.75.-c, 73.22.-f

\vskip 0.6 truecm
\noindent \large{\bf 1. Introduction}

The layered group-IV condensed-matter systems have attracted considerable attention in the fields of physics, materials science and chemistry, mainly owing to the nano-scaled thickness and hexagonal symmetry [1-11]. They have high potentials for the near-future technological applications, e.g., nano-electronics [12,13], optoelectronics [14,15] and energy storage [16,17]. Graphene [18], silicene [19-21], germanene [22-24] and tinene [25] have been successfully synthesized on distinct substrates, such as, C on SiC surface [18], Si on Ag(111), Ir(111) and Zr$B_{2}$ surfaces [19-21], Ge on Pt(111), Au(111) and Al(111) surfaces [22-24], and Sn on B$\mathrm{i}_{2}$T$\mathrm{e}_{3}$ surface [25]. Monolayer graphene exhibits a planar structure with strong $\sigma$ bondings. The others have low-buckled structures arising from the competition of  $sp^2$ and $sp^3$ bondings; furthermore, they possess the significant spin-orbital coupling (SOC) [1-6]. The SOC will play an important role in low-lying energy bands of Si, Ge and Sn. The essential physical properties can be easily tuned by changing the dimensionality and applying the magnetic and electric fields. A 1D nanoribbon could be regarded as a 2D layer cut along the longitudinal direction. This work is focused on the feature-rich electronic properties of 1D nanoribbons, especially for the unique magnetic quantization. Germanene nanoribbons are chosen as a model study because of the low-energy ($\pi$, $\pi^*$) bands and the non-negligible SOC. The dependence on the finite-size effect, the SOC and the external fields is investigated in detail. A detailed comparison with graphene nanoribbons is also made.

A lot of theoretical [11,26-30] and experimental [31-35] studies have been done for the electronic properties of graphene nanoribbons.
The 1D energy bands are mainly determined by the finite width (W) and the edge structure, e.g., the W-dependent energy gap ($E_g$)
in armchair systems and the partial flat bands in zigzag systems [27,28,30]. They could be dramatically changed by a uniform perpendicular magnetic field ($B_z\hat z$), while the magnetic length is comparable to the ribbon width. The competition between the quantum confinement and the magnetic quantization can create the coexistent quasi-Landau levels (QLLs) and  parabolic dispersions in band structures [9,27]. The electronic properties of germanene nanoribbons are expected to be greatly diversified by the buckled structure and the SOC; that is,
there exist certain important differences between germanene and graphene nanoribbons.

The tight-binding model, with the 4$p_z$ orbitals, is used to investigate the low-energy electronic properties of germanene nanoribbons. The effects due to the complicated relations among the quantum confinement, the SOC, and the magnetic and electric fields are explored in detail. This work shows that the unique electronic properties are revealed in energy dispersions, energy gaps, state degeneracy, spatial charge distributions, and spin configurations. Such features are quite different between germanene and graphene nanoribbons. Moreover, the four mechanisms can create three kinds of wave functions and two types of spin states. The rich electronic energy spectra are directly reflected in many special structures of density of states (DOS). They could be examined by the experimental measurements of scanning tunneling spectroscopy (STS) [36-40].

\vskip 0.6 truecm
\noindent \large{\bf 2. The Peierls tight-binding model}

  A zigzag germanene ribbon (ZGR), as shown in Figs. 1(a) and 1(b), is chosen for a model study. A ZGR has two sublattices composed of Ge atoms at A and B sites, respectively. The  sublattice distance in the buclked structure is 2$\ell$=0.66 ${\AA}$. The period of the lattice along the $x$-axis is $I_{x}$=4.02 ${\AA}$, and the first Brillouin zone is defined by $-1\leq k_{x}\leq1$ in the unit of $\pi/I_{x}$. The ribbon width is characterized by the number of zigzag lines along the y-direction, and a primitive unit cell has $2N_{y}$ Ge atoms. The low-energy physical properties, even with the SOC, are dominated the 4$p_z$ orbitals, e.g., the $\pi$-electronic structure. The Hamiltonian built from the $4p_z$-orbital tight-binding functions is given by [3,4]

\begin{eqnarray}
\begin{array}{l}
H=-\sum\limits_{\langle i,j\rangle _{\alpha }}\gamma _{0}c_{i\alpha
}^{+}c_{j\alpha }+i\frac{\lambda _{so}}{3\sqrt{3}}\sum\limits_{\langle
\langle i,j\rangle \rangle _{\alpha \beta }}\nu _{i,j}c_{i\alpha }^{+}\sigma
_{\alpha \beta }^{z}c_{j\beta }\\
+i\frac{2\lambda _{R}}{3}\sum\limits_{\langle \langle i,j\rangle \rangle
_{\alpha \beta }}\mu _{i,j}c_{i\alpha }^{+}(\overrightarrow{\sigma }\times
\hat{d_{ij}})_{\alpha \beta }^{z}c_{j\beta }+\sum\limits_{i_{\alpha
}}U_{i}c_{i_{\alpha }}^{+}c_{i_{\alpha }},
\end{array}
\end{eqnarray}

where $c_{i\alpha}^+$ ($c_{j\alpha}$) is a creation (annihilation) operator of an electron with spin polarization $\alpha$ at the i site. The parameters $\gamma_{0}$=1.04 eV, $\lambda_{so}$=43 meV, and $\lambda_{R}$=10.7 meV [3] are associated with the nearest-neighbor atomic interaction, the effective spin-orbit coupling, and the intrinsic Rashba SOC, respectively. The sum considers all pairs of the nearest neighbor ($\langle i,j\rangle$) and next-nearest neighbor ($\langle\langle i,j\rangle\rangle$). The first term in Eq. (1) is kinetic energy. The second term represents the effective SOC, where $\overrightarrow{\sigma}$= ($\sigma_{x}$, $\sigma_{y}$, $\sigma_{z}$) is the vector of Pauli matrix, with $\nu_{i,j}$=+1 ($-$1) for the anticlockwise (clockwise) next-nearest-neighbor interaction as referring to the direction of +$\hat{z}$. $\hat{d_{ij}}$ in the third term is an unit vector connecting the same sublattice at the i and j sites, and $\mu_{i,j}$=+1 ($-$1) is used for sublattice A (B). In the fourth term, $U_{i}$= +e$E_{z}\ell$ ($-$e$E_{z}\ell$) is the Coulomb potential energy of sublattice at A (B) site, owing to a perpendicular external electric field $E_{z}\hat{z}$. The Bloch wave function is expressed as:


\begin{eqnarray}
\left| \psi^{c,v} \right\rangle  = \sum\limits_{m = 1}^{_{^{{N_y}}}} {a_m^\uparrow} \left| {A_m^\uparrow} \right\rangle  + b_m^\uparrow\left| {B_m^\uparrow} \right\rangle  + {a_m^\downarrow} \left| {A_m^\downarrow} \right\rangle  + b_m^\downarrow\left| {B_m^\downarrow} \right\rangle ,
\end{eqnarray}

where $c$ and $v$, respectively, correspond to the conduction and valence states. $|A_{m}^{\uparrow,\downarrow}\rangle$ ($|B_{m}^{\uparrow,\downarrow}\rangle$) is the tight-binding function associated with the periodic  $A_{m}$ ($B_{m}$) atom with a specific spin configuration. The superscripts, $\uparrow$ and $\downarrow$, represent the atoms in the spin-up and spin-down states, respectively. When a ZGR is subjected to $\textbf{B}$=$B_{z}\hat{z}$, an extra Peierls phase $\Delta G_{ij}$ characterized by the vector potential $\mathbf{A}$=$-B_{z}y\hat{x} $ is introduced in the Hamiltonian matrix elements between the sites $\mathbf{R}_{i}$ and $\mathbf{R}_{j}$ . The hopping parameter $\gamma_{0}$ in Eq. $(1)$ is thus transformed into $\gamma_{0}(B_{z})=\gamma_{0}\exp i[\Delta G_{ij}]$.
The Peierls phase $\Delta G_{ij}$ takes the form of $\frac{2\pi}{\phi_{0}}\int_{\mathbf{R}_{i}}^{\mathbf{R}_{j}}{\mathbf{A}}\cdot d{\mathbf{l}}$, where flux quantum $\phi_{0}=\frac{hc}{e}$.
The Hermitian magnetic Hamiltonian matrix built from the subspaces spanned by the tight-binding functions in the sequence of
$\{|A_{1}^{\uparrow}\rangle,|B_{1}^{\uparrow}\rangle,|A_{1}^{\downarrow}\rangle,|B_{1}^{\downarrow}\rangle,\cdot\cdot\cdot,|A_{m}^{\uparrow}\rangle,|B_{m}^{\uparrow}\rangle,|A_{m}^{\downarrow}\rangle,|B_{m}^{\downarrow}\rangle,\cdot\cdot\cdot,|A_{N_{y}}^{\uparrow}\rangle,|B_{N_{y}}^{\uparrow}\rangle,|A_{N_{y}}^{\downarrow}\rangle;|B_{N_{y}}^{\downarrow}\rangle\}$ has a band-like form [10]

  %

\begin{eqnarray}
H = \left[ {\begin{array}{*{20}{c}}
   {\textbf{H}_{\mathbf{1,1}} } & {\mathbf{H}_{\mathbf{1,2}} } & {} & {} & {} & {}  \\
   {\textbf{H}_{\mathbf{2,1}} } & {\textbf{H}_{\mathbf{2,2}} } &  \ddots  & {} & {0} & {}  \\
   {} & \begin{array}{l}
  \\
  \ddots  \\
 \end{array} &  \ddots  &  \ddots  & {} & {}  \\
   {} & {} &  \ddots  & {\textbf{H}_{\mathbf{m,m}} } & {\textbf{H}_{\mathbf{m,m+1}} } & {}  \\
   {} & {0} & {} & {\textbf{H}_{\mathbf{m+1,m}} } &  {\textbf{H}_{\mathbf{m+1,m+1}} }  &  \ddots   \\
   {} & {} & {} & {} &  \ddots  &  \ddots   \\
   {} & {} & {} & {} & {} & {}  \\
\end{array}}. \right]
\end{eqnarray}

The Hamiltonian matrix in Eq. (3) is composed of the non-vanishing $N_{y}$ $4\times4$ block matrices, in which the two independent matrices are $\mathbf{H}_{m,m}$ and $\mathbf{H}_{m,m+1}$. Furthermore, all other elements of the Hamiltonian matrix are zeros.
The non-zero matrix elements include

\begin{eqnarray}
\left. \begin{array}{l}
{\left[ {{\mathbf{H}_{m,m}}} \right]_{11}}
\\=2t_{2}\sin\{k_{x}I_{x}\}+eE_{z}\ell \\
{\left[ {{\mathbf{H}_{m,m}}} \right]_{33}}
\\=-2t_{2}\sin\{k_{x}I_{x}\}+eE_{z}\ell \\
{\left[ {{\mathbf{H}_{m,m}}} \right]_{12}}={\left[ {{\mathbf{H}_{m,m}}} \right]_{34}}\\
=2\gamma_{0}\cos\{\frac{k_{x}I_{x}}{2}-\frac{\pi\phi}{\phi_{0}}(m-\frac{N_{y}+1}{2})\} \\
{\left[ {{\mathbf{H}_{m,m}}} \right]_{13}}\\=2it_{1}\sin\{k_{x}I_{x}\}\\
{\left[ {{\mathbf{H}_{m,m}}} \right]_{22}}
\\=-2t_{2}\sin\{k_{x}I_{x}\}-eE_{z}\ell\\
{\left[ {{\mathbf{H}_{m,m}}} \right]_{44}}
\\=2t_{2}\sin\{k_{x}I_{x}\}-eE_{z}\ell\\
{\left[ {{\mathbf{H}_{m,m}}} \right]_{24}}\\=-2it_{1}\sin\{k_{x}I_{x}\} \\ \end{array} \right\} \mathrm{for} 1\leq m\leq N_{y},
\end{eqnarray}

\begin{eqnarray}
\left. \begin{array}{l}
{\left[ {{\mathbf{H}_{m,m+1}}} \right]_{11}}=-{\left[ {{\mathbf{H}_{m,m+1}}} \right]_{33}}
\\=-2t_{2}\sin\{\frac{k_{x}I_{x}}{2}\} \\
{\left[ {{\mathbf{H}_{m,m+1}}} \right]_{13}}=-{\left[ {{\mathbf{H}_{m,m+1}}} \right]_{24}}\\
=-it_{1}[\sqrt{3}\cos\{\frac{k_{x}I_{x}}{2}\}-\sin\{\frac{k_{x}I_{x}}{2}\}] \\
{\left[ {{\mathbf{H}_{m,m+1}}} \right]_{21}}={\left[ {{\mathbf{H}_{m,m+1}}} \right]_{43}}=\gamma_{0}\\
 {\left[ {{\mathbf{H}_{m,m+1}}} \right]_{22}}=-{\left[ {{\mathbf{H}_{m,m+1}}} \right]_{44}}
\\=2t_{2}\sin\{\frac{k_{x}I_{x}}{2}\}\\
{\left[ {{\mathbf{H}_{m,m+1}}} \right]_{31}}=-{\left[ {{\mathbf{H}_{m,m+1}}} \right]_{42}}\\=-it_{1}[\sqrt{3}\cos\{\frac{k_{x}I_{x}}{2}\}+\sin\{\frac{k_{x}I_{x}}{2}\}] \\ \end{array} \right\}{\rm{ for    }} 1\leq m\leq N_{y},
\end{eqnarray}

where the parameters are $t_{2}=\frac{\lambda_{so}}{3\sqrt{3}}$ and $t_{1}=\frac{2\lambda_{R}}{3}$. By solving the Hamiltonian matrix, the energy dispersion $E^{c,v}$ and the wave function $\psi^{c,v}$ are obtained.

\vskip 0.8 truecm
\noindent \large{\bf 3. Magneto-electronic properties}

The electronic structure has the band-edge states situated at K=$k_x$=2/3 and  K$'$=$k_x$=4/3, as shown in Fig. 2(a). That is to say, it exhibits two degenerate valleys. All energy bands have parabolic dispersions except two subbands neatest to the Fermi level ($E_F$=0). Without the SOC, the energy spacing two neighboring subbands decreases with the increase of state energy, owing to the quantum-confinement effect. Specifically, the $n^{c}=0$ and $n^{v}=0$ subbands are dispersionless and degenerate at $E_F$. Two partially flat bands are composed of the localized edge states in the range of $\frac{2}{3}<k_{x}<1$. On the other hand, the band structure is drastically changed by the SOC, as indicated in Fig. 2(b).
The subband spacing has no simple relation with $n^{c}$ and $n^{v}$. When the state energy increases, the subband spacing first grows, and then declines at higher energy. The $n^{c}=0$ and $n^{v}=0$ subbands do not merge together within a certain range of $k_x$; furthermore, they vary from the partially flat bands to the slightly distorted linear bands intersecting at $k_{x}=1$.

If a ZGR is subjected to a uniform perpendicular magnetic field, the electronic states with close energies will flock together. Whether the quasi-Landau levels (QLLs) come to exist is determined by the competition between the magnetic quantization and the quantum-confinement effect. As to a $N_y$=150 ZGR, the lower-$n^{c,v}$ QLLs are formed in the valence and conduction states at $B_z$=15 T (Figs. 3(a) and 3(b)), since their magnetic lengths are smaller than the ribbon width. With the increment of state energy, the QLLs will disappear gradually, and the quantum confinement becomes dominant. It is also noticed that  the two $n^{c,v}$=o subbands are further  split into the spin-dependent four subbands, namely the
($n_{\downarrow}^c$=0,$n_{\uparrow e}^c$=0,$n_{\downarrow e}^v$=0,$n_{\uparrow}^v$=0) subbands (identified from wave functions in Fig. 6). The subscript e represents the edge state. Such bands have the unusual energy dispersions associated with the regular LL states or the localized edge states. Moreover, they determine a small direct gap $E_{g}$=6.3 meV near the zone boundary (Fig. 3(b)), depending on the Rashba SOC.

When a perpendicular electric field is applied, $E_{z}$ can create the spin- and valley-dependent electronic  states simultaneously. The main reason is that the z=0  mirror symmetry is destroyed by the Coulomb potential differences on the A and B sublattices. The spin-up- and the spin-down-dominated QLLs, as indicated in Fig. 4, are different from each other, being denoted by $n^{c,v}_{\uparrow}$ and $n^{c,v}_{\downarrow}$, respectively. Their energy spacing between the $n^{c,v}_{\uparrow}$ and $n^{c,v}_{\downarrow}$ QLLs is sufficiently large in the magnitude of $\sim$20-30 meV, and it is larger for the lower-energy QLLs. Specifically, when the valleys are interchanged between the K and K$'$ points, the spin-up states become the spin-down states, or vice versa. The degenerately valley-dependent states could be  destroyed by $E_z$, especially for the lower-$n^{c,v}$ states (e.g., the $n^{c,v}\le\,1$ states). The magneto-electronic energy spectrum of a buckled ZGR is asymmetric about $k_x=0$ in the presence of $E_z$, since the asymmetry of $x\to\,-x$  is generated by that of $z\to\,-z$.  As to the K (K$'$) valley, the $ n_{\downarrow}^c$=0 and $n_{\uparrow}^v$=0 QLLs (the $n_{\downarrow e}^c$=0 and $n_{\uparrow e}^v$=0 edge states) belong to the occupied states, while the opposite is true for the $n_{\uparrow e}^c$=0 and $n_{\downarrow e}^v$=0 edge states
(the $n_{\uparrow}^c$=0 and $n_{\downarrow}^v$=0 QLLs). Energy gap is almost zero, reflecting the very close energy between the highest occupied $n_{\downarrow}^c$=0 QLL and the lowest unoccupied $n_{\downarrow}^v$=0 QLL (DOS in Fig. 8).

The quantum-confinement effects result in the regular standing waves in a finite-width
ZGR. The spatial probability distributions are clearly shown in Fig. 5 for the low-lying band-edge states at the K valley. They behave like the well-defined standing waves except that the $n^c$=0 state  presents the quick decrease from one edge to another one. The $n^c$=1, 2; 3 states, respectively, possess the 3/4, 5/4; 7/4 wavelengths, regardless of the A or B sublattice; that is, the wavelength is (2$n^c$+1)/4 for the $n^c\,\le\,1$ subband.
The similar wave functions could also be found in valence bands. The above-mentioned features are independent of the spin configuration.

In the presence of $B_{z}$, the standing waves are changed into the symmetric Landau wave functions, except for the edge states, as shown in Figs. 6(a) and 6(b). All the localized wave functions are very sensitive to the spin configurations. The $n^c_{\uparrow e}$=0 state at $k_{x}=2/3$ exhibits the extremely large localization distributions in a certain edge of the $A_{m}^{\uparrow}$ sublattice (Fig. 6(a)), but very small ones in the other sublattices. The similar behavior is revealed in the $n^v_{\downarrow e}$=0 state under the interchange of $A_{m}^{\uparrow}$ and $A_{m}^{\downarrow}$. All QLLs are well-behaved in the spatial distribution. For the $n^{c,v}$=0, 1, 2 and 3 QLL states, they, respectively, possess the 0, 1, 2, and 3 zero points in the dominating B sublattice. However, there are $n^{c,v}-1$ zero points in the A sublattice as a result of the hexagonal symmetry. Each QLL state is composed of two distinct spin configurations; furthermore the spin-up- and spin-down- dominated wave functions are similar to each other.

A perpendicular electric filed causes most of the QLL distribution probabilities to transfer between A and B sublattices, as indicated in Figs. 7(a) and 7(b). The $n^c_{\uparrow e}$=0 and $n^v_{\downarrow e}$=0 edge states and
the $n^c_{\downarrow}$=0 and $n^v_{\uparrow}$=0 QLLs do not alter the spatial probability distribution, while the $n^{c,v}\geq$1 QLLs exhibit the opposite behavior. For the conduction QLLs of $ n^c\ge\,1$, the state probability of the major components is transferred from B sublattice to A sublattice (comparison of Fig. 7(a) and 6(a)). Moreover, the minor spin-down components of $A_{m}^{\downarrow}$ and $B_{m}^{\downarrow}$ sublattice in the spin-up-dependent QLL state almost vanish and vice versa under the interchange of spin configuration. This clearly indicates that the combination of spin-up and spin-down configurations is separated by an electric field, i.e., the QLLs exhibit the spin-decomposed configurations in a composite electric and magnetic field.
In addition, the valence states present the similar probibility transfer in the reversed direction from A to B sublattice (Figs. 7(b) and 6(b)).

Germanene and graphene nanoribbons are very different from each other in electronic properties. As to the former,
the SOC can generate the distorted linear subbands  nearest to $E_F$. These two subbands are further split into four  spin-dependent ones  by the cooperation of SOC and $B_z$.  All QLLs become the spin-dependent ones in a composite $B_z$ and $E_z$; furthermore, the  magneto-electronic spectrum is asymmetric about $k_x$=0, depending on the strength of $E_z$. The above-mentioned features are absent in the latter and will induce more special structures in DOS (Figs. 8(a)-8(c)).

 The main features of electronic structures are directly reflected in DOS. It has a lot of special structures due to the complicated relations among quantum confinement, SOC, magnetic and electric fields. Without the external fields (Fig. 8(a)), the first factor leads to many prominent asymmetric peaks arising from 1D parabolic bands, and the second one results in two deformed shoulder structures near $E_{F}$ associated with distorted linear valence and conduction bands (Fig. 2(b)). DOS is finite in the range of $-$50 meV$\leq\omega\leq$50 meV, clearly indicating the metallic behavior. The magnetic quantization causes the lower-energy  asymmetric peaks to change into the delta-function-like symmetric peaks except for a pair of asymmetric ones nearest to $E_{F}$, as shown in Fig. 8(b). The semiconducting property is evidenced by a small gap arising from the Rashba SOC (Fig. 3(b)). Apparently, the peak structures almost become double in the presence of electric field (Fig. 8(c)), mainly owing to the spin-split QLLs (Fig. 4). Energy gap vanishes after a broadening factor of ($\sim$3 meV) is taken into account, and a very strong peak due to the $n^{c,v}$=0 QLLs is situated near $E_{F}$. Specially, a symmetric peak is, respectively, accompanied with two and one asymmetric peaks at $\omega\sim\,0$ and $\sim\,90$  meV (open blue circles), revealing the valley-dependent energy bands (Fig. 4). The above-mentioned features in DOS can be verified by the STS measurements.

STS is a powerful method in examining the form, energy, number and intensity of special structures
in DOS. The differential tunneling conductance (dI/dV) is approximately proportional
to DOS and can directly present the main structures. The STS measurements have  been successfully utilized
to investigate the diverse electronic properties of the carbon-related
systems, such as, graphene nanoribbons [36], carbon nanotubes [37], few-layer graphenes [38,39], and graphites [40]. For example, the $E_z$-induced energy gaps in few-layer graphenes [38],  the monolayer- and bilayer-like $B_z$-dependent LL spectra in graphene systems [39], and the Landau subbands in graphites [40]
are confirmed by STS. The predicted characteristics of DOS in ZGRs,
the asymmetric peaks, the deformed shoulder structures, the delta-function-like peaks, the spin-split QLL peaks and the neighboring symmetric and asymmetric peaks, could be further verified with STS. Such measurements are useful in understanding the competitive or cooperative relations among the critical four factors, and the differences between germanene and graphene nanoribbons.

\vskip 0.8 truecm
\noindent \large{\bf 4. Concluding Remarks}

Electronic properties of zigzag germanene nanoribbons are studied by using the tight-binding model. They are enriched by the complex relations among the finite-width confinement, the SOC, the magnetic quantization, and the electric field. These mechanisms determine quantum number, energy dispersion, energy gap, state degeneracy, wave function, and spin configuration. There are three kinds of spatial charge distributions, namely the normal standing wave, the well-behaved LL distribution, and the edge-localized one. Furthermore, the spin states include the spin-decomposed configuration, and the up- and down-dominated combination ones. The rich electronic structures are directly reflected in DOS with many special structures. The predicted electronic energy spectra could be verified by the STS measurements.

The distinct mechanisms result in the diverse electronic properties. The quantum confinement causes germanene nanoribbons to exhibit 1D parabolic bands except a pair of partially flat bands nearest to $E_F$ coming from the zigzag boundary. These two kinds of energy bands, respectively, correspond to the regular standing waves and the edge-localized distributions. With the SOC, the metallic behavior is evidenced by the distorted linear bands, and the spin states are changed from the separate configurations into the spin up- and down-dominated ones. The QLLs and unusual energy bands are further created by the magnetic quantization. Furthermore, they become the spin- and valley-dependent electronic states in the presence of electric field, owing to the destruction of z=0 mirror symmetry. The former exhibit the spatial distributions with regular zero points, and the spin-dependent weights are modified by the external fields. Specially, the electric field can induce the probability transfer between A and B sublattices with the same spin. The dramatic changes of energy dispersions are clearly indicated by the deformed shoulders and symmetric and anti-symmetric peaks in DOS, such as, the spin-split QLL peaks and the neighboring latter two  in a composite $B_z$ and $E_z$.

\newpage
{\Large\bf References}
\renewcommand{\baselinestretch}{1}
\begin{itemize}

\item[${[1]}$]
C. L. Kane and E. J. Mele,
Phys. Rev. Lett \textbf{95}, 226801 (2005)

\item[${[2]}$]
C. L. Kane and E. J. Mele,
Phys. Rev. Lett \textbf{95}, 146802 (2005)

\item[${[3]}$]
C. Liu, H. Jiang and Y. Yao,
Phys. Rev. B  \textbf{84}, 195430 (2011)

\item[${[4]}$]
M. Ezawa,
New J. Physics \textbf{14}, 033003  (2012)

\item[${[5]}$]
M. Ezawa,
Phys. Rev. Lett \textbf{109}, 055502 (2012)

\item[${[6]}$]
M. Ezawa,
Phys. Rev. Lett \textbf{110}, 026603 (2013)

\item[${[7]}$]
J. H. Wong, B. R. Wu, and M. F. Lin, J. Phys. Chem. C \textbf{116}, 8271 (2012)

\item[${[8]}$]
S. Y. Lin, S. L. Chang, F. L. Shyu, and M. F. Lin,
Carbon \textbf{86}, 207 (2015)

\item[${[9]}$]
C. Y. Lin, J. Y. Wu, Y. H. Chiu and M. F. Lin, Phys. Chem. Chem. Phys \textbf{17}, 26008 (2015)

\item[${[10]}$]
Y. H. Lai, J. H. Ho, C. P. Chang, and M. F. Lin,
Phys. Rev. B \textbf{77}, 235409 (2008)

\item[${[11]}$]
A. H. Castro Neto, F. Guinea, N. M. R. Perez, K. S. Novoselov, and A. K. Geim, Rev. Mod. Phys \textbf{81}, 109 (2009)

\item[${[12]}$]
R. Qin, C. H. Wang, W. Zhu, and Y. Zhang,
AIP Advances \textbf{2}, 022159 (2012)

\item[${[13]}$]
Z. Ni, Q. Lin, K. Tang, J. Zhang, J. Zhou, R. Qin, Z. Gao, D. Yu and J. Lu, Nano Lett \textbf{12}, 113 (2012)

\item[${[14]}$]
H. Houssa, E. Scalise, K. Sankaran, P. Pourtois, V. V. Afanasev, and A. Stesmans,
Appl. Phys. Lett \textbf{98}, 223107 (2011)

\item[${[15]}$]
B. Huang, H. Deng, H. Lee, M. Yoon, B. G. Sumpter, F. Liu, S. C. Smith and S. Wei, Phys. Rev. X \textbf{4}, 021029 (2014)

\item[${[16]}$]
T. Hussain, S. Chakraborty, A. D. Sarkar, B. Johansson, and R. Ahuja,
Appl. Phys. Lett \textbf{105}, 123903 (2014)

\item[${[17]}$]
G. A. Tritsaris, E. Kaxiras, S. Meng, and E. Wang,
Nano Lett \textbf{13}, 2258 (2013)

\item[${[18]}$]
K. V. Emtsev, A. Bostiwick, K. Horn, J. Jobst, G.L. Kellog, L. Ley, J. L. McChesney,
T. Ohta, S. A. Reshanov, J. Rohrl, E. Rotenberg, A. K. Schmid, D. Waldmann, H. B. Weber, T. Seyller, Nature. Mater \textbf{8}, 203 (2009)

\item[${[19]}$]
L. Tao, E. Cinquanta, D. Chiappe, C. Grazianetti, M. Fanculli, M. Dubey, A. Molle, and D. Akinwande,
Nat. Nanotech. \textbf{10}, 227 (2015)

\item[${[20]}$]
P. Vogt, P. D. Padova, C. Quaresima, J. Avila, E. Frantzeskakis, M. C. Asensio, A. Resta, B. Ealet, and G. L. Lay,
Phys. Rev. Lett. \textbf{108}, 155501 (2012)

\item[${[21]}$]
B. Aufray, A. Kara, S. Vizzini, H. Oughaddou, C. Landri, B. Ealet, and G. L. Lay,
Appl. Phy. Lett. \textbf{96}, 183102 (2010)

\item[${[22]}$]
L. F. Li, S. Z. Lu, J. B. Pan, Z. H. Qin, Y. Q. Wang, Y. L. Wang, G. Y. Cao,
S. X. Du, and J. H. Gao, Adv. Mater. \textbf{26}, 4820 (2014)

\item[${[23]}$]
M. Derivaz, D. Dentel, R. Stephan, M. C. Hanf, A. Mehdaoui, P. Sonnet, and C. Pirri,
Nano Lett. \textbf{15}, 2510 (2015)

\item[${[24]}$]
M. E. D¡¦avila, L. Xian, S. Cahangirov, A. Rubio, and G. L. Lay,
New J. Phys. \textbf{16}, 095002 (2014)

\item[${[25]}$]
F. F. Zhu, W. J. Chen, Y. Xu, C. L. Gao, D. D. Guan, C. H. Liu, D. Qian, S. C. Zhang, and J. F. Jia,
Nat. Mater. \textbf{14}, 1020 (2015)

\item[${[26]}$]
S. L. Chang, B. R. Wu, J. H. Wong, and M. F. Lin,
Carbon \textbf{77}, 1031 (2014)

\item[${[27]}$]
H. C. Chung, C. P. Chang, C. Y. Lin, and M. F. Lin,
Phys. Chem. Chem. Phys \textbf{18}, 7573 (2016)

\item[${[28]}$]
K. Nakada, M. Fujita, G. Dresselhaus, and M. S. Dresselhaus,
Phys. Rev. B \textbf{54}, 17954 (1996)

\item[${[29]}$]
M. Fujita, K. Wakabayashi, and K. Kusakabe,
J. Phys. Soc. Jpn \textbf{65}, 1920 (1996)

\item[${[30]}$]
M. F. Lin, and F. L. Shyu,
J. Phys. Soc. Jpn \textbf{69}, 3529 (2000)

\item[${[31]}$]
M. Y. Han, B. Oezyilmaz, Y. Zhang, and P. Kim,
Phys. Rev. Lett \textbf{98}, 206805 (2007)

\item[${[32]}$]
J. Bai, X. Duan, and Y. Huang,
Nano Lett \textbf{9}, 2083 (2009)

\item[${[33]}$]
C. P. Puls, N. E. Staley, and Y. Liu,
Phys. Rev. B \textbf{79}, 235415 (2009)

\item[${[34]}$]
X. Li, X. Wang, L. Zhang, S. Lee, and H. Dai,
Science \textbf{319}, 1229 (2008)

\item[${[35]}$]
C. Casiraghi, A. Hartschun, H. Qian, S. Piscanec, C. Georgi, A. Fasoli, K. S. Novoselov, D. M. Basko and A. C. Ferrari,
Nano Lett \textbf{9}, 1433 (2009)

\item[${[36]}$]
L. Tapaszto, G. Dobric, D. Lambin, and L. P. Biro,
Nature Nanotech \textbf{3}, 397 (2008)

\item[${[37]}$]
L. C. Venema, J. W. Janssen, M. R. Buitelaar, J. W. G. Wildoer, S. G. Lemay, L. P. Kouwenhoven and C. Dekker,
Phys. Rev. B \textbf{62}, 5238 (2000)

\item[${[38]}$]
R. Xu, L. J. Yin, J. B. Qiao, K. K. Bai, J. C. Nie and L. He,
Phys. Rev. B \textbf{91}, 035410 (2015)

\item[${[39]}$]
A. Luican, G. Li, A. Reina, J. Kong, R. R. Nair, K. S. Novoselov, A. K. Geim and E. Y. Andrei,
Phys. Rev. Lett \textbf{106}, 126802 (2011)

\item[${[40]}$]
M. L. Sadowski, G. Martinez, M. Potemski, G. Berger and W. A. de Heer,
Phys. Rev. Lett \textbf{97}, 266405 (2006)



\end{itemize}

\newpage
{\Large\bf Figure captions}
\renewcommand{\baselinestretch}{1}
\begin{itemize}


\item[Figure 1:] Geometric structure of a zigzag germanene ribbon: (a) top view, and (b) side view


\item[Figure 2:] The band structures, (a) without the SOC effect and external fields; (b) without any fields


\item[Figure 3:] The magnet-electronic structures at $B_z$=15 T, (a) without and (b) with the Rashiba SOC.

\item[Figure 4:] The magneto-electronic structure at a composite field of $B_z$=15 T and $E_z$=0.14 V/$\AA$.

\item[Figure 5:] The spatial probability distributions of the conduction-band states at the K valley of $k_x$=2/3.

\item[Figure 6:] Same plot as Fig. 5, but shown at $B_z$=15 T for (a) conduction and (b) valence states.


\item[Figure 7:] Same plot as Fig. 5, but shown at a composite field of $B_z$=15 T and $E_z$=0.14 V/$\AA$ for (a) conduction and (b) valence states.


\item[Figure 8:] Density of states (a) in the absence of field, and for (b) $B_z$=15 T; (c) $B_z$=15 T and $E_z$=0.14 V/${\AA}$.



\end{itemize}

\end{document}